\newif\ifconfver
\confvertrue        

\ifconfver
	\documentclass{article}
	\usepackage{spconf}
	\title{MULTIUSER ONE-BIT MASSIVE MIMO PRECODING UNDER MPSK SIGNALING }
	\name{Mingjie Shao$^{\dagger}$, Qiang Li$^{\star}$, Yatao Liu$^{\dagger}$ and Wing-Kin Ma$^{\dagger}$
    \thanks{Mingjie Shao's work was supported by Hong Kong PhD Fellowship. Qiang Li's work was supported in part by the National Natural Science Foundation of China under Grant
61531009, and in part by the Fundamental Research Funds for
the Central Universities under Grant ZYGX2016J011.} }
	\address{
	\normalsize{
	$^{\dagger}$ Department of  Elec. Eng., The Chinese University of Hong Kong, Hong Kong SAR, China	} \\
	\normalsize{
	$^{\star}$ School of Info. \& Comm. Eng., University of Electronic Science and Technology of China, China}\\
	\small {E-mail: $^{\dagger}$ \{mjshao, ytliu, wkma\}@ee.cuhk.edu.hk, $^{\star}$  lq@uestc.edu.cn }
	}
\else
	\documentclass[11pt,draftcls,onecolumn]{IEEEtran}
	\usepackage{fullpage}
	\title{ONE-BIT PRECODING FOR MULTIUSER MASSIVE MIMO WITH PSK SIGNALING}
	
\fi
\usepackage{amsmath,graphicx}

\usepackage{calc,amsfonts,amssymb,bm,url,color,theorem,cite}
\usepackage{psfrag,subfigure,float}
\usepackage{algorithm}
\usepackage{algorithmic}
\usepackage{mathtools,lipsum,cuted,multicol}
\usepackage{filecontents}
\usepackage{epstopdf}
\usepackage{array}
\usepackage{caption}

\definecolor{orange}{RGB}{255,107,0}

\newtheorem{Prop}{Proposition}

\newtheorem{Corollary}{Corollary}

\theorembodyfont{\rmfamily}

\newcommand\bv{\ensuremath{{\bm v}}}
\newcommand\bw{\ensuremath{{\bm w}}}

\newcommand\bh{\ensuremath{{\bm h}}}

\newcommand\bu{\ensuremath{{\bm u}}}

\newcommand{\setX}{\mathcal{X}}

\newcommand{\Rbb}{\mathbb{R}}
\newcommand{\Cbb}{\mathbb{C}}

\newcommand\bx{\ensuremath{{\bm x}}}

\newcommand\bb{\ensuremath{{\bm b}}}

\newcommand\bbx{\ensuremath{\bar{\bm x}}}

\newcommand{\setS}{\mathcal{S}}

\newcommand\br{\ensuremath{{\bm r}}}

\newcommand{\Rfrak}{\mathfrak{R}}
\newcommand{\Ifrak}{\mathfrak{I}}

\newcolumntype{M}[1]{>{\centering\arraybackslash}m{#1}}

\makeatletter
\def\bstctlcite{\@ifnextchar[{\@bstctlcite}{\@bstctlcite[@auxout]}}
\def\@bstctlcite[#1]#2{\@bsphack
  \@for\@citeb:=#2\do{%
    \edef\@citeb{\expandafter\@firstofone\@citeb}%
    \if@filesw\immediate\write\csname #1\endcsname{\string\citation{\@citeb}}\fi}%
  \@esphack}
\makeatother

\begin{document}
\ninept
\bibliographystyle{IEEEtran}


\bstctlcite{IEEEexample:BSTcontrol}
\maketitle
\begin{abstract}
Most recently, there has been a flurry of research activities on studying how massive MIMO precoding should be designed when the digital-to-analog conversion at the transmitter side is operated by cheap one-bit digital-to-analog converters (DACs). Such research is motivated by the desire to substantially cut down the hardware cost and power consumption of the radio-frequency chain, which is unaffordable in massive MIMO if high-resolution DACs are still used. One-bit MIMO precoding design problems are much harder to solve than their high-resolution DAC counterparts. In our previous work, we developed a minimum symbol-error probability (SEP) design for one-bit precoding under the multiuser MISO downlink scenario and under quadrature amplitude modulation signaling.
Leveraging on the previous work, this work shows how the minimum SEP design is applied to $M$-ary phase shift keying (MPSK) signaling.
Simulation results show that our minimum SEP design delivers significantly better bit-error rate (BER) performance than the other designs for higher-order PSK such as $8$-PSK and $16$-PSK.
As a minor, but useful, side contribution, we also tackle an MPSK SEP characterization problem which was only intuitively treated in the prior arts.
\end{abstract}


\section{Introduction}

Massive MIMO is a promising physical-layer technique for future wireless communication systems.
It promises high spectral efficiency, robustness against channel fading, and many other good properties~\cite{Rusek2013,Lu2014}.
However, the benefits of massive MIMO come at a price of scaling up the radio-frequency (RF) chains, which have non-negligible hardware costs and are power hungry if high-resolution digital-to-analog converters (DACs)/analog-to-digital converters (ADCs) are employed and good linear dynamic ranges are desired.
 As such, there has been growing interest in implementing massive MIMO by low-cost and power-efficient hardwares, and the use of one-bit ADCs and DACs---which can be cheaply implemented and do not require the RF chains to have high linear dynamic ranges---is seen as a promising solution~\cite{Risi2014,Guerreiro2016}.



The use of one-bit ADCs in massive MIMO was first considered in uplink problems such as symbol detection and channel estimation from one-bit quantized measurements \cite{Choi2015,Choi2016,Studer2016}.
Recently, the research focus moves from one-bit ADC uplink to one-bit DAC downlink.
The work~\cite{Saxena2016} analyzed the performance of one-bit quantized zero-forcing (ZF) precoding when the ratio of the number of transmit antennas to the number of users is high.
Later, in \cite{Swindlehurst2017}, it was shown that adding random perturbations on the symbols is effective in mitigating the quantization errors.
Departing from the notion of designing a linear precoder and then quantizing it,
some works proceed with a direct one-bit precoder design approach.
In \cite{Jacobsson2017} and \cite{Jacobsson2017a}, minimum mean square error (MMSE)-based one-bit precoding schemes were proposed to minimize the distance between the desired symbols and the transmit symbols for single-carrier and OFDM systems, respectively.
Some other designs exploit the underlying symbol constellation structures for enhancing the symbol-error probability performance.
The work \cite{Jedda2017} dealt with the $M$-ary phase shift keying (MPSK) case, and developed a precoding solution based on linear program (LP) relaxation;
see also \cite{Masouros2015} and \cite{Amadori2017} which described similar ideas in the contexts of constructive interference and constant envelope precoding, respectively.
In our recent work \cite{Shao2018}, we considered a minimum symbol-error probability (SEP) design for the quadratic amplitude modulation (QAM) case.
There, we developed a non-convex optimization approach to handle the very difficult nature of the one-bit precoding problem, namely, the binary constraint arising from the restriction of one-bit signal transmission.
Numerical results showed promising performance with our design.

The goal of this paper is to apply our non-convex optimization approach to the MPSK case.
The details with the approach we take on will become clear later as we proceed to the main development.
As will be shown numerically, the application of our approach to the MPSK case shows superior bit-error rate (BER) performance compared to the existing (and very recently developed) one-bit precoder designs for MPSK.
Also, as a minor, but useful, side contribution, we address an MPSK SEP characterization problem which was intutively treated in the previous studies.
Again, this will become clear as we describe the problem in the next section.

 \section{Problem Formulation}
 \label{sec:format}


Our one-bit massive MIMO precoding problem is described as follows.
Our scenario of interest is that of a multiuser downlink, where a base station (BS), equipped with $N$ transmit antennas, transmits information signals to $K$ single-antenna users in a simultaneous and unicast fashion.
We follow the widely-used system model in this context, where the relationship of the transmitted and received signals is modeled as
 \begin{equation} \label{model}
 \begin{aligned}
 y_{i,t}= \bh_{i}^T \bx_t + \eta_{i,t} ,~~i=1,\ldots, K,~ t=1,\ldots, T.
 \end{aligned}
 \end{equation}
 Here, $\bx_t \in \Cbb^N$ is the multi-antenna signal transmitted by the BS at symbol time $t$;
 $y_{i,t}$ is the signal received by user $i$ at symbol time $t$;
 $\bh_i \in \Cbb^N$ is the channel associated with user $i$;
 $\eta_{i,t}$ is noise and is assumed to be circular complex Gaussian with mean zero and variance $\sigma^2$;
 $T$ is the length of the transmission block.
The BS employs a massive antenna array, implemented by one-bit DACs.
This leads to the restriction
\[
 \bx_t \in {\cal X} \triangleq
 \left \{ \pm \sqrt{\tfrac{P}{2N}} \pm \sqrt{\tfrac{P}{2N}} j \right\}^N,
 \]
 where $P$ is the total transmit power.

Our precoding problem is to design $\{ \bx_t \}_t$ such that every user will receive its own symbol stream, with the symbol-error probability being as small as possible.
To put into context, let $\{ s_{i,t} \}_t$ be the symbol stream for user $i$.
In this work, we assume the MPSK constellation where
\begin{equation} \label{Sconstel}
s_{i,t} \in {\cal S }\triangleq \{s~|~s=  e^{jn\frac{2\pi}{M}},~n=0,\ldots,M-1\}.
\end{equation}
Also, let ${\rm dec}: \Cbb \rightarrow \setS$ be the MPSK decision function, i.e., ${\rm dec}(y) = e^{j \hat{n} \frac{2\pi}{M}}$ where $\hat{n} \in \{ 0,\ldots,M-1 \}$ is such that the phase angle of $y$ lies in $[ \frac{2\pi \hat{n}}{M} - \frac{\pi}{M}, \frac{2\pi \hat{n}}{M} + \frac{\pi}{M}  ]$.
At the users' side, each user detects their symbol stream by $\hat{s}_{i,t} = {\rm dec}(y_{i,t})$.
Let
\begin{equation} \label{SEP}
{\sf SEP}_{i,t} = {\rm Pr}( \hat{s}_{i,t} \neq s_{i,t} | s_{i,t} )
\end{equation}
denote the symbol error probability (SEP) of detecting $s_{i,t}$ conditioned on $s_{i,t}$.
The problem is to find an appropriate $\bx_t$ such that all ${\sf SEP}_{i,t}$'s will be as small as possible.
Specifically, we consider a precoding design problem
\begin{equation} \label{Pone}
\begin{aligned}
\min_{\bx_t}  ~ \max_{i=1,\ldots,K} {\sf SEP}_{i,t} ~~~~{\rm s.t.}  ~ \bx_t \in {\cal X},
\end{aligned}
\end{equation}
where we seek to minimize the worst user's SEP under the one-bit constraint.

We will focus on how problem \eqref{Pone} is tackled.
But before we proceed to our main development, we should shed some light on the intuitive side of the problem.
The SEP for MPSK symbols, in its exact form, does not admit a simple expression in general.
But suppose that we impose a restriction on $\bx_t$, namely,
\begin{equation} \label{approx}
\bh_i^T \bx_t = \alpha_{i,t} s_{i,t}, \qquad \text{for all $i$},
\end{equation}
for some $\alpha_{i,t} > 0$.
Then, it is well-known in the classical digital communication literature that the SEP admits a simple upper-bound approximation
 \begin{equation} \label{SEP_0}
 {\sf SEP}_{i,t} \leq 2Q \!\left(\!\frac{\alpha_{i,t}}{\sigma\!/\!\sqrt{2}}\!\sin\frac{\pi}{M}\!\right),
 \end{equation}
 where $Q(x)=\int_{x}^{\infty} \frac{1}{\sqrt{2\pi}} e^{-z^2/ 2} dz$;
see \cite[Eqn.~(8.26)]{Simon2005}.
By applying the above approximation to problem \eqref{Pone}, one can readily see that the corresponding objective function can be reduced to $\max_{i=1,\ldots,K} -\alpha_{i,t}$---which is easy to handle.
Unfortunately,
while the restriction \eqref{approx} can be easily satisfied by applying ZF when we do not have the one-bit constraint, it is not clear whether and how \eqref{approx} may be enforced in the presence of the one-bit constraint.
This will lead us to study an SEP approximation that does not require \eqref{approx}.

We should also take this opportunity to mention related works.
It is believed in \cite{Masouros2015,Jedda2017,Swindlehurst2018} that the SEP can be reduced by increasing the so-called safety margin, which is given by
\begin{equation} \label{safe_mag}
\alpha_{i,t}  = \Rfrak\{ \bh_i^T \bx s_{i,t}^{*} \}-|\Ifrak\{ \bh_i^T \bx s_{i,t}^{*} \}| \cot\left(\frac{\pi}{M}\right).
\end{equation}
Readers are referred to the aforementioned references for the intuitions that led to the safety margin.
Note that \eqref{safe_mag} does not require \eqref{approx}.
The approximation \eqref{safe_mag} is simple and greatly simplifies the subsequent MIMO precoding design, as shown in the aforementioned references.
However, up to this point, there has not been a study that mathematically underlies how sound the approximation \eqref{safe_mag} is.

%

\section{The Proposed Minimum SEP Design}

\subsection{SEP Analysis}
\label{sec:SER}

We first address the SEP characterization problem described in the last section.
The problem boils down to a basic probability problem as follows:
We have an observation
\begin{equation}\label{eq:AWGN}
  w=z+\eta,
\end{equation}
where $z \in \Cbb$ can take any value and does not necessarily lie in the symbol constellation ${\cal S}$ in \eqref{Sconstel};
$\eta$ is circular complex Gaussian with mean zero and variance $\sigma^2$.
The problem is to find, through analyses, a tractable approximation of the probability
\[
{\rm Pr}( {\rm dec}(w) \neq 1).
\]

\begin{figure}[htb!]
	\centering
	\includegraphics[width=0.9\linewidth]{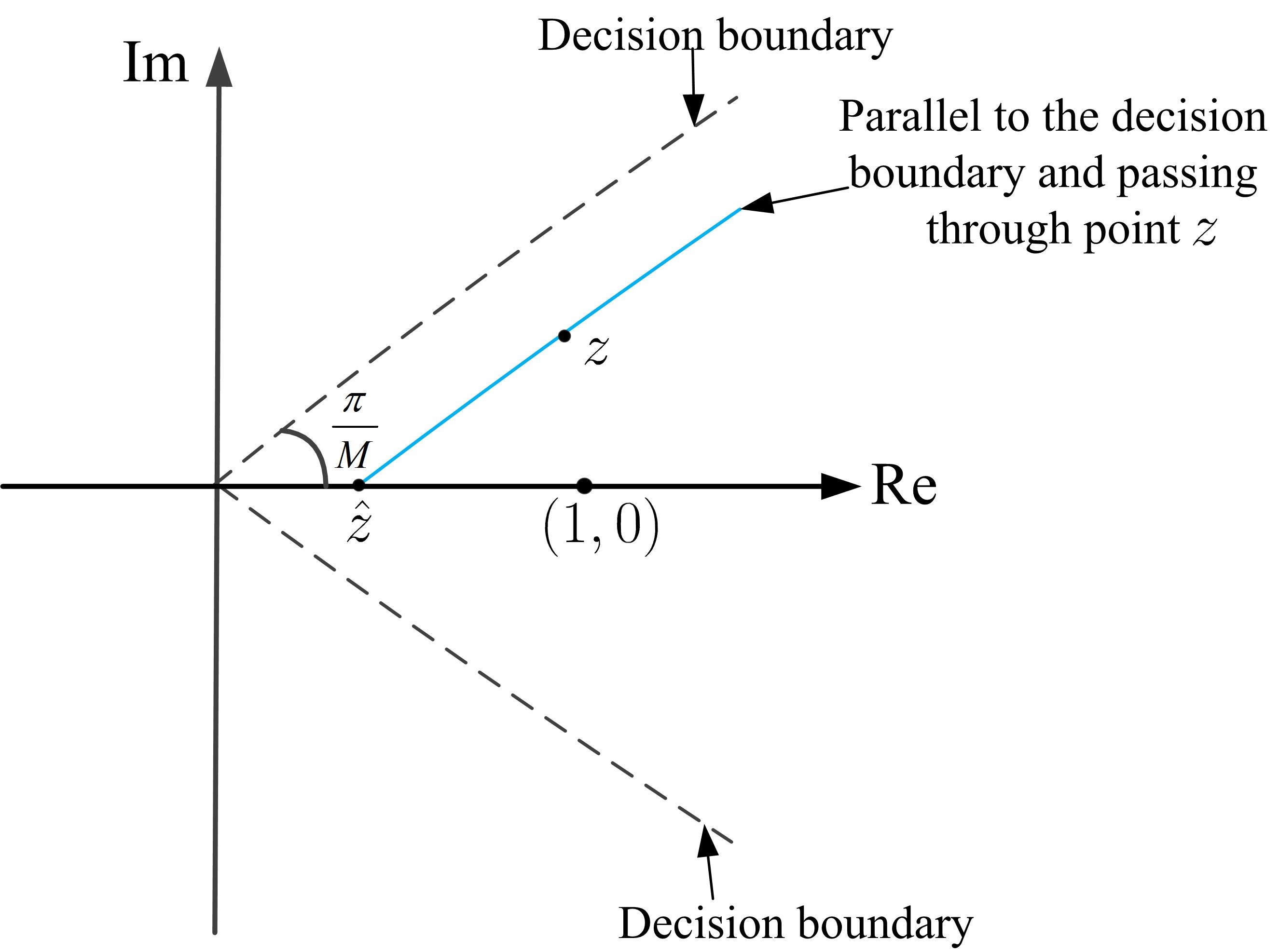}
	\caption{Illustration of $\hat{z}$.}\label{fig:DB}
\end{figure}

We start our analysis with considering a perturbed version of \eqref{eq:AWGN}
\begin{equation}\label{eq:perturb_model}
\hat{w} = \hat{z} + \eta, \quad
\hat{z} = z  + \Delta z,
\end{equation}
where  $\Delta z=-|\Ifrak\{ z\} |\cot(\pi/M) - j \Ifrak\{ z\}$.
One can verify that
\[
\hat{z} = \alpha,
\]
where
\[
\alpha=\Rfrak\{ z\} -|\Ifrak\{ z\}|\cot\left(\frac{\pi}{M} \right).
\]
We should point out that the above $\alpha$ takes the same form as the safe margin in \eqref{safe_mag}.
Fig.~\ref{fig:DB} illustrates how $z$ and $\hat{z}$ are related on the Cartesian plane.
We have the following result.
\begin{Prop}\label{cl:SER_za}
	It holds that	
	\begin{equation} \label{eq:claim_up_bd}
	{\rm Pr}( {\rm dec}(w) \neq 1) \leq {\rm Pr}( {\rm dec}(\hat{w}) \neq 1)
	\end{equation}
	Also, we have
	\begin{equation}\label{eq:claim_up}
	{\rm Pr}( {\rm dec}(\hat{w}) \neq 1) \leq 2 Q \left( \frac{\alpha}{\sigma/\sqrt{2}}\sin\frac{\pi}{M} \right)
	\end{equation}
\end{Prop}
The proof of Proposition~\ref{cl:SER_za} is shown in the Appendix.

As the direct consequence of applying Proposition~\ref{cl:SER_za} to our MIMO precoding problem, we have the following result.
\begin{Corollary} \label{Cor}
	The SEP \eqref{SEP} of the one-bit precoding problem in the last section admits an upper-bound approximation
	\[
	{\sf SEP}_{i,t} \leq 2 Q \left( \frac{\alpha_{i,t}}{\sigma/\sqrt{2}}\sin\frac{\pi}{M} \right)
	\]
	where $\alpha_{i,t}$ is defined as the safety margin in \eqref{safe_mag}.
\end{Corollary}
Corollary~\ref{Cor} shows an interesting revelation: the safety margin used in prior work is sound in that it leads to an upper-bound approximation of the SEP.

\subsection{Reformulation, and the Algorithm}

We are now ready to attack the one-bit precoding design problem in \eqref{Pone}.
By applying Corollary~\ref{Cor} to problem \eqref{Pone}, and using the monotonicity of $Q$, we obtain an approximation of problem \eqref{Pone} as follows
\begin{equation}\label{eq:OB}
\begin{aligned}
\max_{\bx_t} ~\min_{i=1,\ldots, K} \alpha_{i,t}~~~~\text{s.t.} ~ \bx_t \in \setX,
\end{aligned}
\end{equation}
where $\alpha_{i,t}$ is given in \eqref{safe_mag}.
For the sake of notational brevity, let us remove the subscript ``$t$'' in our subsequent expressions.
By complex-to-real conversion,  we reformulate problem~\eqref{eq:OB} as
\begin{equation}\label{eq:OBreal}
\begin{split}
\min_{\bbx} ~~\max_{i=1,\ldots, K} ~\max\{\bu_i^T\bbx,~\bw_i^T\bbx\}~~~~\text{s.t.} ~~ \bbx\in \setX_{\Re},
\end{split}
\end{equation}
where
\begin{equation*}
\begin{split}
\setX_{\Re} & \triangleq \left\{\bx\in \Rbb^{2N}~|~x_i=\pm \sqrt{{P}/{2N}},~~i=1,\ldots,2N\right\},\\
\bbx&=[\Re\{ \bx\}^T, ~\Im\{ \bx\}^T]^T,  ~~~
\bb_i =[\Re\{ s_i^{*}\bh_i^T\}, ~-\Im\{ s_i^{*}\bh_i^T\}]^T,\\
\br_i&= \cot \left(  \frac{\pi}{M} \right) [\Im\{ s_i^{*}\bh_i^T\}, ~~\Re\{ s_i^{*}\bh_i^T\}]^T,\\
\bu_i & =-\bb_i+\br_i,~~\bw_i  =-\bb_i - \br_i.
\end{split}
\end{equation*}

Problem \eqref{eq:OBreal} is a non-convex non-smooth optimization problem.
We tackle it by applying a most recently proposed technique by us \cite{Shao2018}.
The technique has three steps.

\medskip
\noindent{\it \underline{Step 1: Smoothing the objective function}}
\medskip

We apply smooth approximation to the objective function of problem \eqref{eq:OBreal}, thereby circumventing non-smoothness of the problem.
Specifically, we replace the (non-smooth) point-wise maximum function by a (smooth) log-sum-exponential function.
The resulting approximate problem of problem \eqref{eq:OBreal} is given by
\begin{equation}\label{eq:OBsm}
\begin{split}
\min_{\bbx \in \setX_{\Re}} &~ ~f(\bbx)\triangleq \mu \log \sum_{i=1}^K \big(e^{\frac{\bu_i^T\bbx}{\mu}}+e^{\frac{\bw_i^T\bbx}{\mu}}\big)
\end{split}
\end{equation}
where $\mu>0$ is a pre-specified constant that controls the smoothing accuracy. In particular, the approximation is tight as $\mu\rightarrow 0$.

%

\medskip
\noindent{\it \underline{Step 2: Binary constraint reformulation}}
\medskip

We apply a reformulation that will turn the original problem, which is discrete, to a continuous problem.
It can be shown that the following equivalence holds:
\begin{equation*}
	\bm x \in \{ -1,+1 \}^n \Longleftrightarrow \exists \bv: -\mathbf{1}\leq \bx \leq \mathbf{1},~ \| \bv \|_2^2\leq n, ~ \bm x^T \bm v = n.
\end{equation*}
Leveraging the above equivalence, we consider the following reformulation of problem~\eqref{eq:OBsm}
\begin{equation}\label{eq:OBvr}
  \begin{split}
    \min_{\bbx,\bv} &~ ~F_{\lambda}(\bbx,\bv)\triangleq f(\bbx)+ \lambda (P-\bbx^T \bv)\\
    \text{s.t.} &~~ -\sqrt{\frac{P}{2N}} {\bm 1} \leq\bbx \leq \sqrt{\frac{P}{2N}} \mathbf{1}, ~||\bv||_2^2 \leq P,
  \end{split}
\end{equation}
where $\lambda>0$ is a penalty parameter for enforcing  $\bbx^T \bm v = P$.
It can be shown that for a sufficiently large $\lambda$, problem~\eqref{eq:OBvr} is equivalent to problem~\eqref{eq:OBsm}; see \cite[Lemma~1]{Shao2018}.

\medskip
\noindent{\it \underline{Step 3: Alternating minimization}}
\medskip

We apply alternating minimization to problem~\eqref{eq:OBvr}.
Fixing $\bbx$, the minimization with respect to (w.r.t.) $\bm v$ has a closed form
%
\begin{equation}\label{eq:vprob}
\bv=\begin{cases}
\sqrt{P}\bar{\bx}/{\|\bar{\bx}\|_2}~, & ~\mbox{for}~\bbx \neq \bm 0, \\
\mbox{any feasible $\bm v$}, & ~ {\rm otherwise}.
\end{cases}
\end{equation}
Fixing $\bm v$, the minimization w.r.t. $\bar{\bx}$ is a smooth convex optimization problem with (simple) box constraints.
We apply an accelerated projected gradient (APG) algorithm~\cite{IntroCVX,Beck2009} to solve the problem.
Readers are referred to \cite{Shao2018} for the detailed implementations of APG. Algorithm~\ref{Alg:OB} summarizes the overall procedure of handling problem~\eqref{eq:OBreal}.


~\\[-3em]
\begin{algorithm}[htb!]
	\caption{Fast ALternating Minimization (FALM) for~\eqref{eq:OBreal}}
	\begin{algorithmic}[1]
		\STATE Initialize  $\lambda$, $\delta>1$, $\mu$, $\bbx^0=\bv^0=\mathbf{0}$ and iteration index $k=0$
		\REPEAT
		\STATE Fix $\bv = \bv^k$ in \eqref{eq:OBvr} and update $\bar{\bx}^{k+1}$ by the APG algorithm;
		\STATE Update $\bv^{k+1}$ according to \eqref{eq:vprob} with $\bbx = \bbx^{k+1}$;
		\STATE Update $\lambda= \lambda\times \delta$ every $M$ iterations;
		\STATE  $k=k+1$;
		\UNTIL {$\lambda$ is greater than some threshold $\lambda>\lambda_0$.}
	\end{algorithmic}\label{Alg:OB}
\end{algorithm}

~\\[-3em]

\section{Simulation Results and Conclusion}
\label{sec:sim}

We evaluate the performance of our algorithm using Monte-Carlo simulations.
For convenience, we will name our algorithm  (Algorithm~\ref{Alg:OB}) ``FALM''.
We benchmark it with the following algorithms:
zero-forcing (ZF) with infinite-resolution DACs, which will be named ``ZF'';
ZF followed by one-bit quantization, which will be named ``ZF-OB'';
the SQUID algorithm which is based on the MMSE design~\cite{Jacobsson2017};
and the LP relaxation algorithm for maximum safety margin (MSM) design \cite{Jedda2017}, which will be named ``MSM.''


A number of $1,000$ channel realizations were used to test FALM and the benchmarked algorithms.
Each channel realization was randomly generated, following the standard i.i.d. circular complex Gaussian distribution.
The transmission block length is $T= 100$.
The total transmit power is $P=1$.
The number of transmit antennas is $N=128$, and the number of users is $K=24$.

We should also mention the parameter settings of FALM.
The smoothing parameter is $\mu=0.01$.
The penalty parameter $\lambda$ is initialized as $0.01$, and $\delta$ is set as $\delta=10$. Algorithm~\ref{Alg:OB} stops when $\lambda>100$, i.e. we update  $\lambda$ for $5$ times.


\begin{figure}[htb!]
\centering
\includegraphics[width=0.7\linewidth]{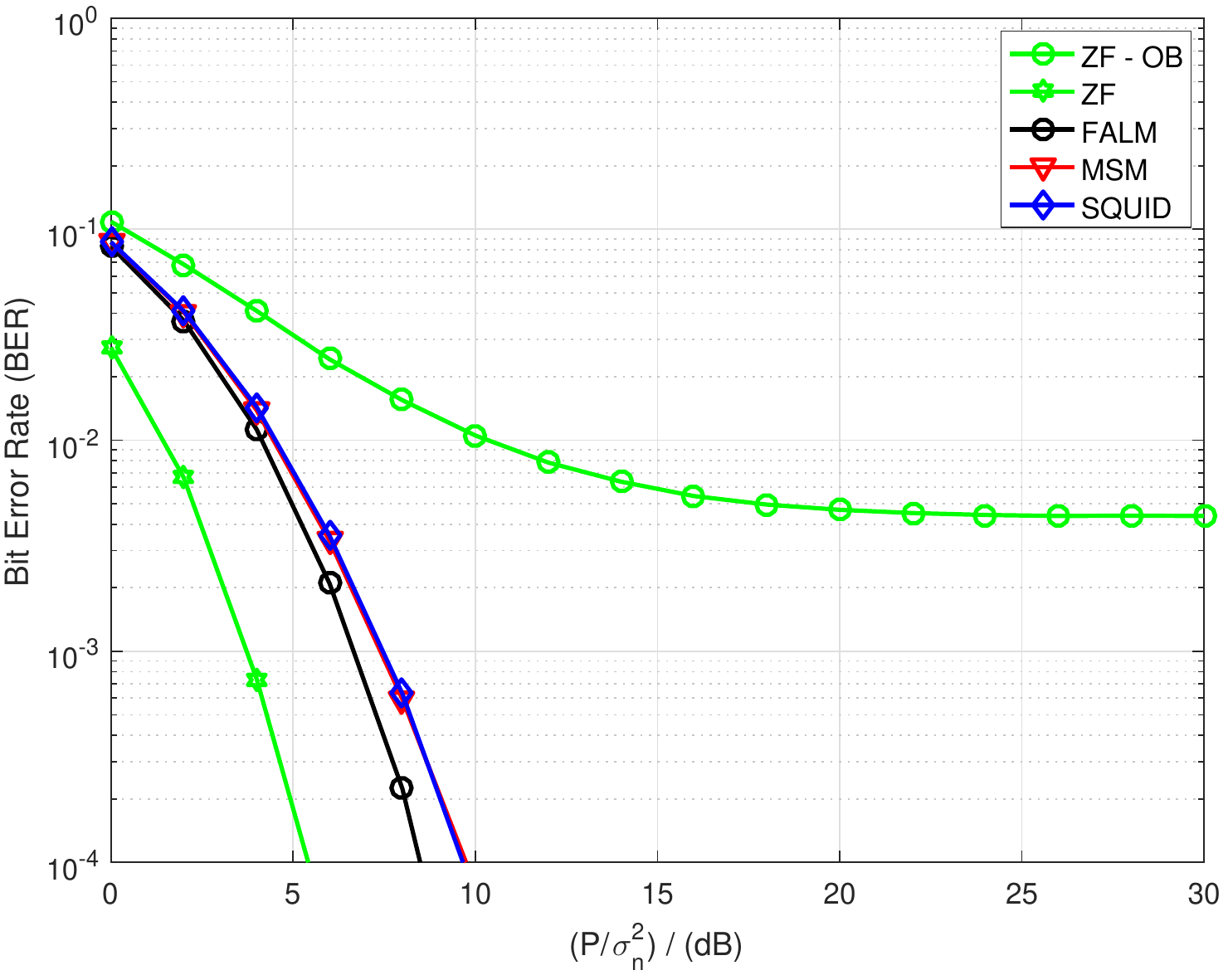}
\caption{Average BER performance versus $P/\sigma_n^2$; QPSK.}\label{fig:4sim}
\end{figure}

\begin{figure}[htb!]
\centering
\includegraphics[width=0.7\linewidth]{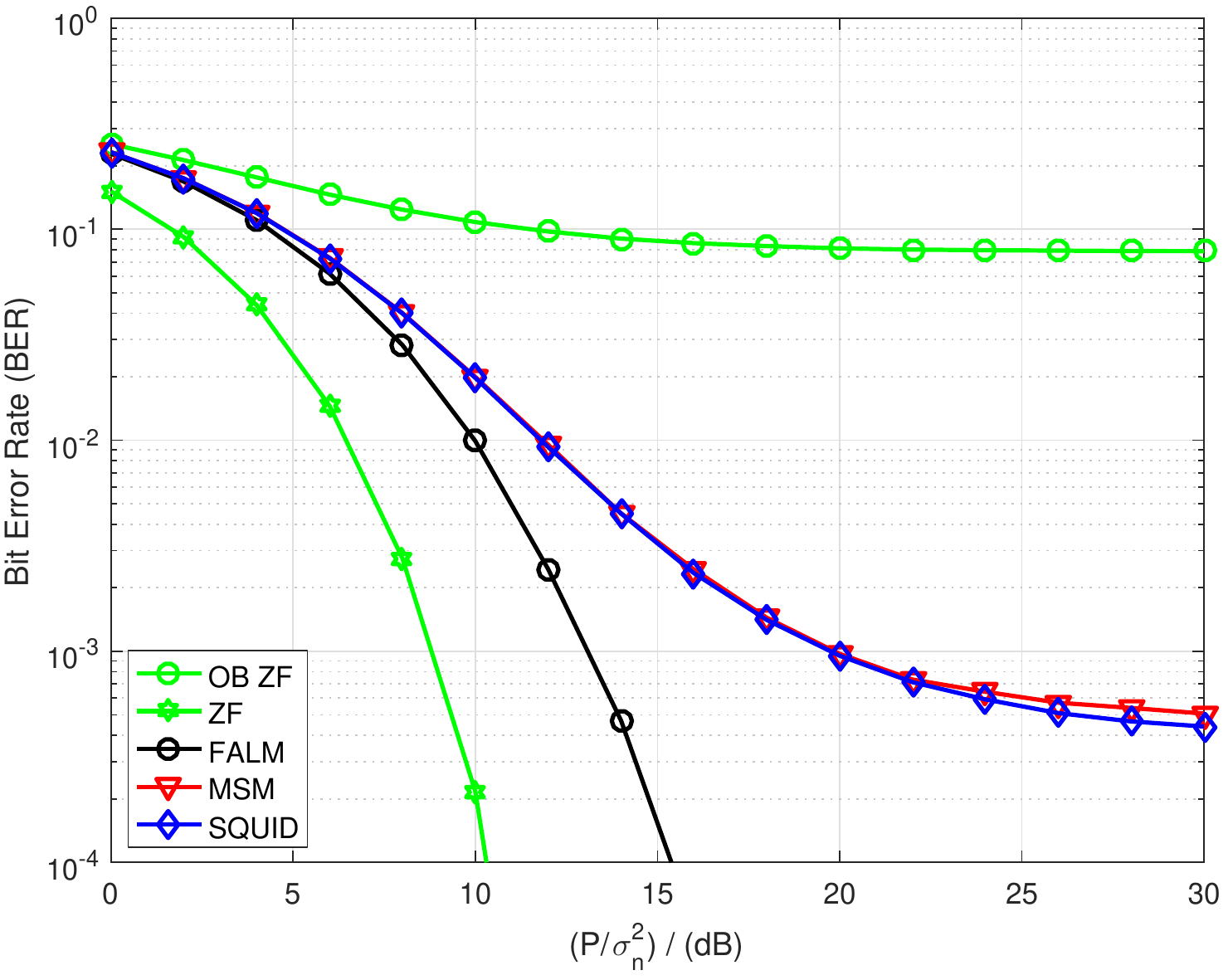}
\caption{Average BER performance versus $P/\sigma_n^2$; $8$-PSK.}\label{fig:8sim}
\end{figure}

\begin{figure}[htb!]
\centering
\includegraphics[width=0.7\linewidth]{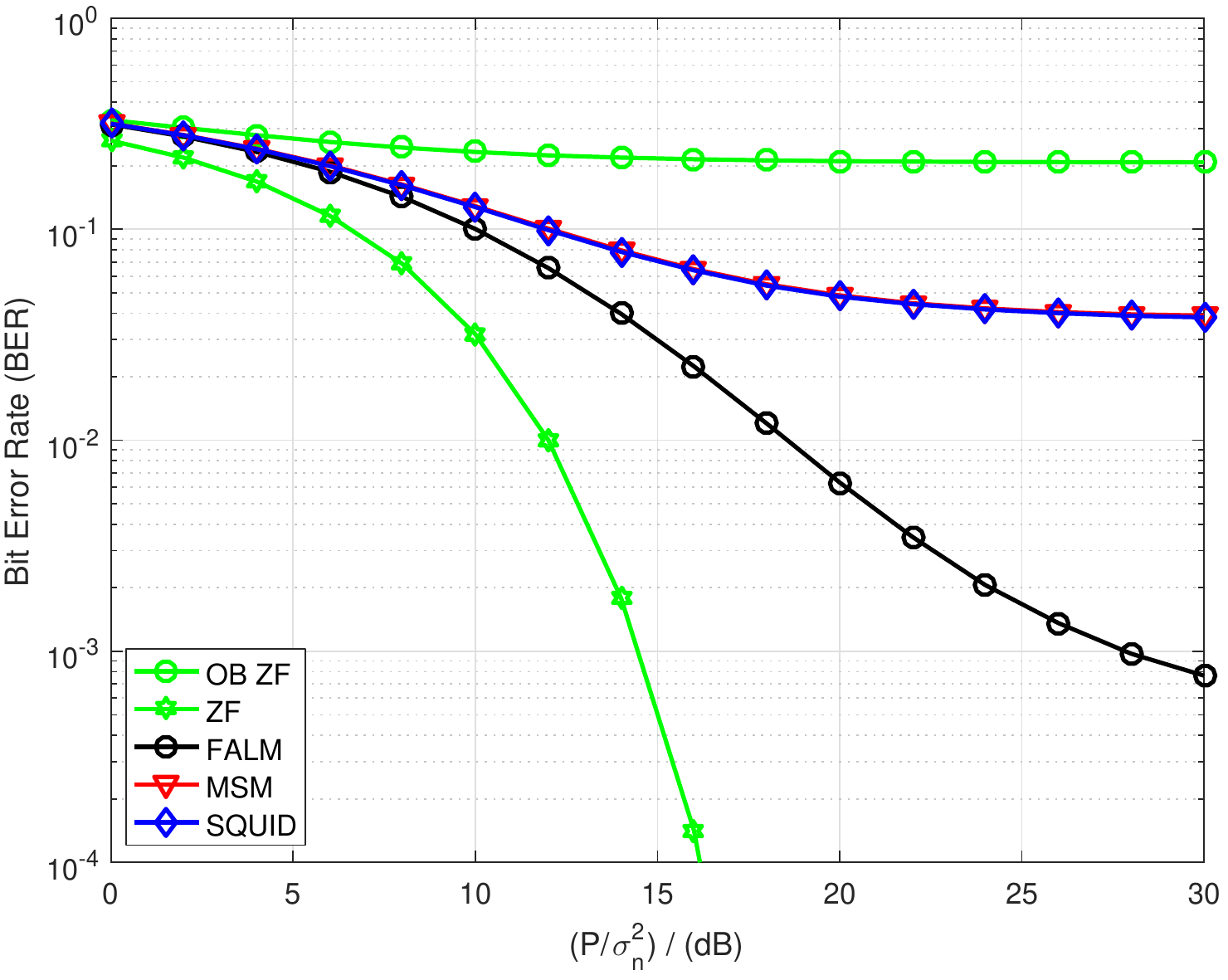}
\caption{Average BER performance versus $P/\sigma_n^2$; $16$-PSK.}\label{fig:16sim}
\end{figure}

Figs. \ref{fig:4sim}--\ref{fig:16sim} show the BER plots of the various algorithms under QPSK, $8$-PSK and $16$-PSK, respectively.
We can see that FALM outperforms the other one-bit precoding algorithms, and the performance gaps are significant for $8$-PSK and $16$-PSK.
In fact, for $8$-PSK and $16$-PSK, SQUID and MSM are seen to suffer from error floor effects.
In comparison, FALM does not  have the same problem.


\begin{figure}[htb!]
\centering
\includegraphics[width=0.7\linewidth]{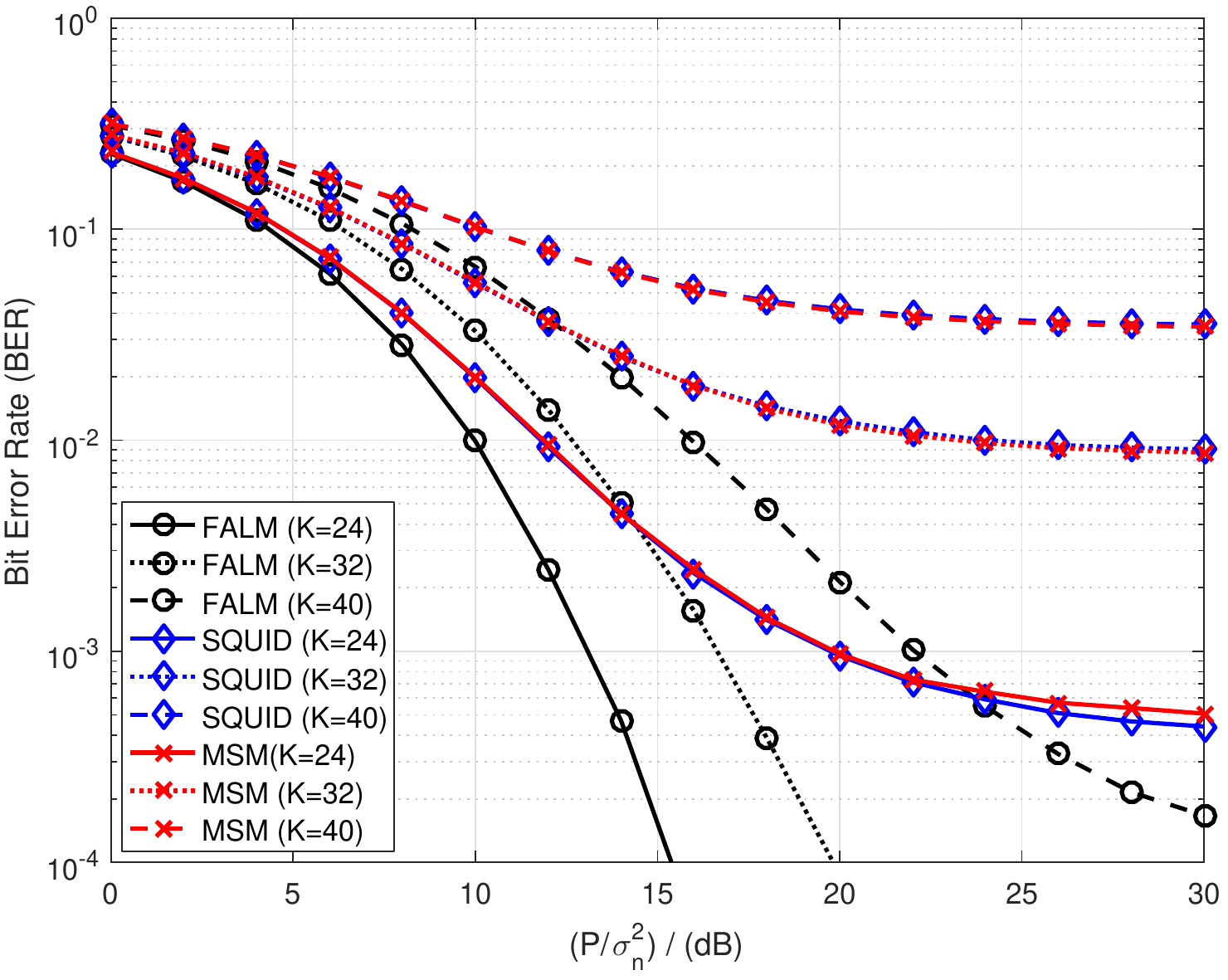}
\caption{Average BER performance for different No. of users; $8$-PSK.}\label{fig:user}
\end{figure}

Fig. \ref{fig:user} evaluates how the BER performance of FALM, SQUID and MSM scales with the number of users.
The number of transmit antennas is again $N=128$;
the MPSK constellation size is fixed at $8$;
the number of users is varied from $K=24$ to $K=40$.
We once again see that FALM outperforms SQUID and MSM.

%


To conclude, we have developed a one-bit precoding algorithm for multiuser MISO downlink and under MPSK symbol constellations.
The algorithm is based on a minimum symbol-error probability formulation, and it applies a continuous non-convex optimization methodology to tackle the very difficult binary constraint in one-bit precoding.
Numerical results show that the proposed algorithm yields superior BER performance.

%


\section{Appendix}\label{appendix:claim1}
Let us first prove the inequality \eqref{eq:claim_up_bd}. It suffices to show that for any given noise realization $\eta$, the following implication holds true:
\begin{equation} \label{eq:claim1_proof}
\text{dec}(\hat{w}) = 1 ~~\Longrightarrow~~ \text{dec}(w) = 1.
\end{equation}
In words, if the noisy reception $\hat{w}$ can be correctly decoded as $1$, then so does $w$. To this end, we notice that the left-hand side of \eqref{eq:claim1_proof} is equivalent to
\begin{subequations}
	\begin{align}
	\Re\{z\}-|\Im\{z\}|\cot(\pi/M)+\Re\{\eta\}\geq & 0,\label{eq:a1}\\
	\frac{\Im\{\eta\}}{\Re\{z\}-|\Im\{z\}|\cot(\pi/M)+\Re\{\eta\}}\in & [-\tan(\pi/M), \tan(\pi/M)].\label{eq:a2}
	\end{align}
\end{subequations}
From \eqref{eq:a1}, we have
\begin{equation}\label{eq:a3}
\Re\{z\} + \Re\{\eta\} \geq |\Im\{z\}| \cot(\pi/M) \geq 0,
\end{equation}
and from \eqref{eq:a2}, we have
\begin{equation}\label{eq:a4}
\begin{split}
\Im\{\eta\} \geq-(\Re\{z\} + \Re\{\eta\})\tan(\pi/M) + |\Im\{z\}|, \\
\Im\{\eta\} \leq (\Re\{z\} + \Re\{\eta\})\tan(\pi/M) - |\Im\{z\}|.
 \end{split}
\end{equation}
The first inequality in \eqref{eq:a4} implies
\begin{equation}\label{eq:a5}
-\tan(\pi/M)  \leq \frac{\Im\{\eta\} - |\Im\{z\}|}{ \Re\{z\} + \Re\{\eta\}} \leq  \frac{\Im\{\eta\} +\Im\{z\}}{ \Re\{z\} + \Re\{\eta\}},
\end{equation}
where the second inequality in \eqref{eq:a5} is due to $\Im\{\eta\} - |\Im\{z\}| \leq \Im\{\eta\}  + \Im\{z\}$ and \eqref{eq:a3}. Similarly, from the second inequality in \eqref{eq:a4}, we have
\begin{equation}\label{eq:a6}
\frac{\Im\{\eta\} + \Im\{z\}}{\Re\{z\}+\Re\{\eta\}} \leq 	\frac{\Im\{\eta\} + |\Im\{z\}|}{\Re\{z\}+\Re\{\eta\}}  \leq \tan(\pi/M).
\end{equation}
Combining \eqref{eq:a3}, \eqref{eq:a5} and \eqref{eq:a6}, we get
\begin{align}
\Re\{z\}+\Re\{\eta\}\geq & 0,\\
\frac{\Im\{z\} + \Im\{\eta\}}{\Re\{z\} +\Re\{\eta\}}\in & [-\tan(\pi/M), \tan(\pi/M)],
\end{align}
which means that $ w = z+ \eta$ lies in the correct decision region of symbol $1$, i.e, $\text{dec}(w) = 1$.

Next, we prove the inequality \eqref{eq:claim_up}. If $\alpha \geq 0$, the upper bound \eqref{SEP_0} directly holds from the well-known union bound for MPSK \cite{Simon2005}:
\begin{equation*}
  \begin{split}
    \Pr(\text{dec}(\hat{w})\neq 1)
    \leq 2Q\left(\frac{\alpha}{\sigma/\sqrt{2}}\sin\frac{\pi}{M}\right).
  \end{split}
\end{equation*}
 If $\alpha<0$, we can apply the similar techniques as in the derivation of upper bound \cite{Simon2005} to obtain
\begin{equation*}
  \begin{split}
    \Pr(\text{dec}(\hat{w})\neq 1)
    \leq & 2\left( 1- Q\left(\frac{-\alpha}{\sigma/\sqrt{2}}\sin\frac{\pi}{M}\right)\right)\\
    =&2Q\left(\frac{\alpha}{\sigma/\sqrt{2}}\sin\frac{\pi}{M}\right),
  \end{split}
\end{equation*}
where the equality is due to the fact $Q(x)=1-Q(-x)$.
This completes the proof.

\vfill\pagebreak
\newpage

\bibliographystyle{IEEEtran}
\bibliography{refs}
\nocite{*}

\end{document}